\documentstyle[12pt]{article}
\begin{document}

\title{Interaction between Noncommutative Open String and Closed-String Tachyon}

\author{
  Akira Kokado\thanks{E-mail: kokado@kobe-kiu.ac.jp},\  
  Gaku Konisi\thanks{E-mail: konisi@kwansei.ac.jp} \  
  and \ 
  Takesi Saito\thanks{E-mail: tsaito@yukawa.kyoto-u.ac.jp} \\
  \\
 {\small{\it{Kobe International University, Kobe 655-0004, Japan${}^{\ast }$}}} \\
 {\small{\it{Department of Physics, Kwansei Gakuin University, Nishinomiya 662-8501, 
 Japan${}^{\dagger ,\ddagger }$}}}}

\date{\small{March 2001}}
\maketitle

\begin{abstract}
 We construct a vertex operator which describes an emission of the ground-state tachyon of the closed string out of the noncommutative open string.  Such a vertex operator is shown to exist only when the momentum of the closed-string tachyon is subject to some constraints coming from the background $B$  field. The vertex operator has a multiplicative coupling constant $g(\sigma )$ which depends on $\sigma $ as $g(\sigma )=\sin^2 \sigma $ in $0 \le \sigma \le \pi $. This behavior is the same as in the ordinary  $B=0$ case.
 
\end{abstract}

\setlength{\parindent}{1cm}
\newpage

\renewcommand{\thesection}{\Roman{section}.}
\renewcommand{\theequation}{\arabic{section}.\arabic{equation}}
\setcounter{equation}{0}

\section{Introduction}
\indent

 The concept of a quantized spacetime was first proposed by Snyder [1], and has received much attention over the past few years [2-6]. Especially interesting is a model of open strings propagating in a constant antisymmetric $B$ field background. Previous studies show that this model is related to the noncommutativity of D-branes [7,8], and in the zero slope limit to noncommutative gauge theories [5]. \\
\indent 
     In the present paper we would like to construct a vertex operator which describes an emission of the ground-state tachyon of the closed string out of the noncommutative open string. Such a vertex operator has been considered in the ordinary open string theory [9], and for the emission of a graviton in the literatures [10],[11]. Let us call such a tachyon the closed-string tachyon. Contrary to the open-string tachyon, the closed-string tachyon can be emitted from any point of the open string. The vertex operator contains a multiplicative coupling constant $g(\sigma )$, which depends on $\sigma $ as $g(\sigma )=\sin^2 \sigma $ in  $0 \le \sigma \le \pi $. When the constant antisymmetric $B$ field is present, the open string becomes noncommutative at both end-points. In this case we will find that such a vertex operator exists only when the momentum of the closed-string tachyon is subject to some constraints coming from the background $B$ field. \\
\indent
     In Sec.II the model of noncommutative open string is summarized. In Sec.III we construct the vertex operator for the closed-string tachyon. The final section is devoted to concluding remarks.

\section{Noncommutative open string}
\indent

 In order to fix notations let us summarize the model of bosonic open string propagating in a constant antisymmetric $B$ field background.  In the conformal gauge the world sheet action of the open string is
\begin{equation}
\label{201}
 S=-\frac{1}{4\pi \alpha '}\int _\Sigma d^2 \sigma [\eta _{\mu \nu }\partial _\alpha X^\mu  \partial ^\alpha X^\nu  - \epsilon ^{\alpha \beta } B_{\mu \nu }\partial _\alpha X^\mu  \partial _\beta X^\nu ],
\end{equation}
where $\eta _{00}=-1$,$\Sigma $ is an oriented world-sheet with boundary (signature (-1, 1)) and $B_{\mu \nu }$ is assumed to be constant.  For simplicity we have set $2\alpha '=1$.  The equation of motion and boundary conditions follow from this action
\begin{equation}
\label{202}
 \partial _\alpha \partial ^\alpha X^\mu  = 0, 
\end{equation}
\begin{equation} 
\label{203}
 \eta _{\mu \nu }X'^\nu + B_{\mu \nu } \dot{X}^\nu |_{\sigma =0,\pi }=0.
\end{equation}
We are looking at a D25-brane with the constant $B$ field.  From (\ref{202}) and (\ref{203}) one obtains the following solution
\begin{equation}
\label{204}
 X^\mu (\tau ,\sigma ) = q^\mu + a_0^\mu \tau  + (\frac{\pi }{2}-\sigma )B^\mu _{\ \nu }a_0^\nu 
 + \sum _{n\neq 0}\frac{1}{n}e^{-in\tau }(ia_n^\mu \cos n\sigma - B^\mu _{\ \nu }a_n^\nu \sin n\sigma ).
\end{equation}
The conjugate momentum is given by
\begin{eqnarray}
\label{205}
 P_\mu &=& \frac{1}{\pi }(\eta _{\mu \nu }\dot{X}^\nu + B_{\mu \nu }X'^\nu ) \nonumber \\
        &=& \frac{1}{\pi }\sum _{n=-\infty }^{\infty }G_{\mu \nu }a_{n}^{\nu }e^{-in\tau }
        \cos n\sigma ,
\end{eqnarray}
where
\begin{equation}
\label{206}
 G_{\mu \nu } = \eta _{\mu \nu } - B_{\mu }^{\ \rho }B_{\rho \nu }.
\end{equation}
According to the Dirac quantization for this constrained system[12],[13], we obtain the following commutation relations:
\[
 [X^\mu (\tau ,\sigma ), P_\nu (\tau ,\sigma ')] = i\delta ^{\mu }_{\ \nu }
 \delta _{c}(\sigma ,\sigma ') 
\]
\begin{equation}
\label{209}
 [P_\mu (\tau ,\sigma ), P_\nu (\tau ,\sigma ')] = 0,
\end{equation}
\[ 
 [X^\mu (\tau ,\sigma ), X^\nu (\tau ,\sigma ')] = i\pi \theta ^{\mu \nu } 
 \{ 1 - \epsilon (\sigma + \sigma ')\},
\]
where $\epsilon $ is the sign function, and the noncommutative parameter $\theta ^{\mu \nu }$  is defined as
\begin{equation}
\label{210}
 \theta ^{\mu \nu } = -B^{\mu }_{\ \rho }(G^{-1})^{\rho \nu }.
\end{equation}
From (\ref{209}) one finds commutators for normal modes
\[
 [a_{m}^{\mu }, a_{n}^{\nu }] = m\delta _{m+n,0}(G^{-1})^{\mu \nu }, 
\]
\begin{equation}
\label{211}
 [q^{\mu }, a_{n}^{\nu }] = i\delta _{n,0}(G^{-1})^{\mu \nu }, 
\end{equation}
\[
  [q^{\mu }, q^{\nu }] = 0.  
\]
\indent

 Now let us write (2.4) as follows\footnote{There is an ambiguity in the zero mode when $X(\tau ,\sigma )$ is divided into $X_{+}$ and $X_{-}$. However, there causes no effect in the result.}:
\[
 X(\tau ,\sigma ) = \frac{1}{2}[X_{+}(\tau + \sigma ) + X_{-}(\tau - \sigma )],  
\]
\begin{eqnarray}
\label{212}
 X_{+} &=& q + (a_0 - Ba_0)(\tau + \sigma ) + \frac{\pi }{2}Ba_0 
 + i\sum _{n=0}\frac{1}{n}e^{-in(\tau + \sigma )}(a_n - Ba_n),  \\
 X_{-} &=& q + (a_0 + Ba_0)(\tau - \sigma ) + \frac{\pi }{2}Ba_0 
 + i\sum _{n=0}\frac{1}{n}e^{-in(\tau - \sigma )}(a_n + Ba_n). \nonumber 
\end{eqnarray}
In the following we use a complex number $z=e^{\tau + i\sigma }$. By the replacement $\tau \to -i\tau $ in (\ref{212}) we have
\[
 X_{+}(z) = q + \frac{\pi }{2}\sinh \beta \cdot \alpha _0 - ie^{-\beta }\alpha _0 \ln z 
 + i\sum _{n\neq 0}\frac{1}{n}z^{-n}e^{-\beta }\alpha _{n},
\]  
\begin{equation}
\label{213}
 X_{-}(\bar{z}) = q + \frac{\pi }{2}\sinh \beta \cdot \alpha _0 - ie^{\beta }\alpha _0 \ln \bar{z} 
 + i\sum _{n\neq 0}\frac{1}{n}\bar{z}^{-n}e^{\beta }\alpha _{n}, 
\end{equation}
where $\beta $ and $\alpha _n$ are defined by
\[
 B = \tanh \beta,
\] 
\begin{equation}
\label{214}
 a_n = \cosh \beta \cdot \alpha _n .
\end{equation}
The "metric" $G_{\mu \nu }$ in (2.6) is related to the "vielbein" $\xi =\cosh ^{-1}\beta $ through equations
\begin{equation}
\label{215}
 G = 1-B^2 = 1 - \tanh ^2 \beta = \cosh ^{-2} \beta = \xi \xi .
\end{equation}
Commutation relations for $\alpha _n$ and $q$ are
\[
 [\alpha _m, \alpha _n] = m\delta _{m+n,0}, 
\]
\begin{equation}
\label{216}
 [q, \alpha _0] = i\cosh \beta .
\end{equation}
The Virasoro operator is then given by the energy-momentum tensor $T_{\pm }(z)$
\[
 L_n = \frac{1}{2\pi i}\oint dzz^{n+1} T_{\pm }(z) = \frac{1}{2}\sum :\alpha _k \alpha _{n-k}:,
\]
\begin{equation}
\label{217}
 T_{\pm }(z) = \frac{1}{2}:J_{\pm }(z)^2:, 
\end{equation}
\[
 J_{\pm }(z) = i\partial _z X_{\pm }(z) = \sum e^{\mp \beta }\alpha _n z^{-n-1} 
\]
and satisfies the same Virasoro algebra as in the ordinary open string theory without the $B$ field.

\section{Vertex operator for closed string tachyon}
\setcounter{equation}{0}
\indent

 Let $V(z, \bar{z})$ be the vertex operator which describes the emission of the closed string tachyon out of the noncommutative open string. The Virasoro operator $\tilde{L}(f)$ of this interacting system is defined as 
\begin{equation}
\label{301}
 \tilde{L}(f) = \frac{1}{\pi }\int _0^\pi d\sigma (\tilde{T}_{00}f^0 + \tilde{T}_{01}f^1),
\end{equation}
where $\tilde{T}_{\alpha \beta }$ is the energy-momentum tensor with
\begin{equation}
\label{302}
 \tilde{T}_{00} = \frac{1}{2}:(\dot{X}^2 + X'^2): + V, \qquad
 \tilde{T}_{01} = :\dot{X}X':,
\end{equation} 
and $f^\alpha $ is expressed by an arbitrary function $f(\tau \pm \sigma )$ as
\begin{equation}
\label{303}
 f^0 = \frac{1}{2}[f(\tau + \sigma ) + f(\tau - \sigma )] , \qquad
 f^1 = \frac{1}{2}[f(\tau + \sigma ) - f(\tau - \sigma )]. 
\end{equation}
All operators are considered in the interaction picture. Equation (3.1) can be rewritten into the free Virasoro operator plus the vertex part
\begin{equation}
\label{304}
 \tilde{L}(f) = L(f) + V(f), 
\end{equation}
where
\[
 L(f)=\frac{1}{2\pi }\int _{-\pi }^{\pi }d\sigma f(\tau + \sigma ):\frac{1}{2}(\dot{X}+X')^2:,  
\]
\begin{equation}
\label{305}
 V(f)=\frac{1}{\pi }\int _{0}^{\pi }d\sigma f^0(\tau ,\sigma )V(\tau ,\sigma ). 
\end{equation}
Note that $V(\tau ,\sigma )$ is not an even function of $\sigma $.  In the complex number
\begin{equation}
\label{306}
 L(f) = \frac{1}{2\pi i}\oint dz zf(z)T_{\pm}(z).
\end{equation}
Here the integration path is a closed path around the origin.  \\
\indent
 The operator $\tilde{L}(f)$ should satisfy the Virasoro algebra
\begin{equation}
\label{307}
 [\tilde{L}(f), \tilde{L}(g)] = i\tilde{L}(f\stackrel{\leftrightarrow}{\partial }g) +  \textrm{anomaly term}.
\end{equation}
This is equivalent to following equations:
\begin{equation}
\label{308}
 [L(f), V(g)] - [L(g), V(f)] = iV(f\stackrel{\leftrightarrow}{\partial }g), 
\end{equation}
\begin{equation}
\label{309}
 [V(f), V(g)] = 0.
\end{equation}
In the following we look for the vertex operator which satisfies Eqs.(\ref{308}) and (\ref{309}). Let us assume
\begin{equation}
\label{310}
 V(z,\bar{z}) = U_{+}(z)U_{-}(\bar{z})
\end{equation}
and
\begin{equation}
\label{311}
 U_{\pm}(w) = :\exp\{ikX_{\pm}(w)\}:.
\end{equation}
We then calculate the operator product expansion of $T_{\pm}(z)$ and $U_{\pm}(w)$.  First we find the contraction
\begin{equation}
\label{312}
 <X_{\pm}(z)X_{\pm}(w)> = -\ln(z-w) - (e^{\mp \beta }\cosh\beta - 1)\ln z + \textrm{const}.
\end{equation}
If we define $\phi _{\pm}(w)$ by $\phi _{\pm}(w)=ikX_{\pm}(w)$, then it follows that
\begin{equation}
\label{313}
 <J_{\pm}(z)\phi _{\pm}(w)> = \frac{1}{z-w}k + (e^{\mp \beta }\cosh\beta - 1)k\frac{1}{z}.
\end{equation}
Since
\begin{equation}
\label{314}
 <J_{\pm}(z)U_{\pm}(w)> = <J_{\pm}(z):\exp\phi _{\pm}(w):> 
 = <J_{\pm}(z)\phi _{\pm}(w)>\frac{\partial U_{\pm}(w)}{\partial \phi _{\pm}(w)},
\end{equation}
one gets, dropping the $\pm$ suffices,
\begin{eqnarray}
 <T(z)U(w)> &=& <\frac{1}{2}:J(z)J(z):U(w)> \nonumber \\ 
\label{315}
 &=&:J(z)<J(z)U(w)>: + \frac{1}{2}J(z)^\bullet J(z)^{\bullet \bullet }U(w)^{\bullet ,\bullet \bullet }
 \\
 &=&i\partial _zX(z)[\frac{1}{z-w}k + (e^{\mp\beta }\cosh\beta - 1)k\frac{1}{z}]\frac{\partial U(w)}{\partial \phi (w)} 
 + \frac{1}{2}[***]^2\frac{\partial ^2U(w)}{\partial \phi^2 (w)}. \nonumber
\end{eqnarray}
The first $1/(z-w)$  term becomes
\begin{equation}
\label{316}
 \partial _z\phi (z)\frac{1}{z-w}\frac{\partial U(w)}{\partial \phi (w)} 
 = \frac{1}{z-w}\partial _wU(w) + \textrm{regular term}.
\end{equation}
The second $1/z$ term is regular around $z=w$. The third bracket squared-term reduces to
\begin{equation}
\label{317}
 [***]^2=[\frac{1}{z-w}k-\bar{\beta }k\frac{1}{z}]^2=\frac{k^2}{(z-w)^2}
 -2\frac{k\bar{\beta }k}{z-w}\frac{1}{z} 
 +\frac{k\bar{\beta }^2k}{z^2},
\end{equation} 
where
\begin{equation}
\label{318}
 \bar{\beta }\equiv 1-e^{\mp \beta }\cosh \beta . 
\end{equation}
To sum up we have the operator product expansion 
\begin{equation}
\label{319}
 T(z)U(w)=\frac{k^2/2}{(z-w)^2}U(w)
 +\frac{1}{z-w}\partial _wU(w)-\frac{k\bar{\beta }k}{z-w}\frac{1}{z}U(w)
 +\textrm{regular term}.
\end{equation}
The third term in the right-hand side violates the conformal covariance of $U(w)$. So, in the following we will put a constraint for the momentum $k$
\begin{equation}
\label{320}
 k\bar{\beta }k = 0.
\end{equation}
Thus we find that the vertex operator $U(w)$  has the conformal weight $h=k^2/2$. The equation (3.19) without the third term is equivalent to the equation
\begin{equation}
\label{321}
 [L(f), U(w)]=(k^2/2)\partial [wf(w)]U(w)+wf(w)\partial U(w).
\end{equation}
When
\begin{equation}
\label{322}
 k^2=2,
\end{equation} 
the right-hand side becomes a total derivative $\partial [wf(w)U(w)]$. Compatibility of (3.22) with (3.20) will be discussed later. In this case one finds the equation
\begin{equation}
\label{323}
 [L(f), V(w,\bar{w})]=\partial _w[wf(w)V(w,\bar{w})]
 + \partial_{\bar{w}}[\bar{w}f(\bar{w})V(w,\bar{w})].
\end{equation}
In the real time formulation the right-hand side can be written as
\begin{eqnarray}
 -i\partial _\alpha [f^\alpha V]+2f^0V &=& -i\partial _0[f^0V]-i\partial _1[f^1V]+2f^0V \nonumber \\
\label{324}
 &=& -i\partial _0f^0V-if^0\partial _0V-i\partial _1[f^1V]+2f^0V \\
 &=& -i\partial _1f^1V-i\partial _0f^0V-i\partial _1[f^1V]+2f^0V, \nonumber
\end{eqnarray}
so that we obtain
\[
 [L(f), V(g)] = \frac{1}{\pi }\int ^\pi _0 d\sigma  g^0\{-i\partial _\sigma f^1V
 -if^0\partial _\tau V-i\partial _\sigma (f^1V)+2f^0V\}.
\]
Hence there holds
\begin{eqnarray}
[L(f), V(g)] - [L(g), V(f)] \nonumber \\
 &=& -i\frac{1}{\pi }\int ^\pi _0 d\sigma g^0[\partial _\sigma f^1V+\partial _\sigma (f^1V)]
 -(f\leftrightarrow g) \nonumber \\
 &=& -i\frac{1}{\pi }\int ^\pi _0 d\sigma [g^0\partial _\sigma f^1-f^1\partial _\sigma g^0]V
 -(f\leftrightarrow g) \nonumber \\
\label{325}
 &=& i\frac{1}{\pi }\int ^\pi _0 d\sigma [f^1\partial _\sigma g^0+f^0\partial _\sigma g^1]V
 -(f\leftrightarrow g) \\
 &=& i\frac{1}{2\pi }\int ^\pi _0 d\sigma [(f\partial g)(\tau +\sigma )
 +(f\partial g)(\tau -\sigma )]V-(f\leftrightarrow g) \nonumber \\
 &=& iV(f\stackrel{\leftrightarrow}{\partial }g). \nonumber
\end{eqnarray}
This assures Eq.(\ref{308}). \\
\indent
 Now let us discuss the compatibility of Eq.(3.20) with Eq.(3.22). The Eq.(3.18) can be rewritten as \begin{equation}
\label{326}
 \bar{\beta } = \pm e^{\mp \beta }\sinh \beta .
\end{equation}
Hence Eq.(3.20) is equivalent to the equation
\begin{equation}
\label{327}
 B_{\mu \nu }k^\nu = 0.
\end{equation}
This gives  $k_i = k_j = 0$ for $i,j$ with $B_{ij}\neq  0$ for the canonical form.  The on-shell condition $k^2 = 2$ holds only with the elements in the subspace $B_{ij}=0$. [A possible four-dimensional example satisfying these two conditions is 
$(k_0, k_1, 0, 0)$  with 
$k_1^{\ 2} - k_0^{\ 2} = 2$  for 
$B_{01}=0$, $B_{23}\neq  0$. The three-dimensional momentum $(k_1, 0, 0)$ turns out to be in the direction of the magnetic field 
$(B_{1}\equiv B_{23}, 0, 0)$.] From the constraints (3.20) and (3.22) one can deduce 
\begin{equation}
\label{328}
 2 =ke^{\pm\beta }\cosh \beta \cdot k = ke^{\pm 2\beta }k = k\cosh 2\beta \cdot k.
\end{equation}

\indent
     Let us then rewrite the vertex operator $V(z,\bar{z})=U_+(z)U_-(\bar{z})$ into the normal product. Making use of the formulas (3.28) we have
\begin{eqnarray}
 V(z,\bar{z})&=&:e^{ikX_{+}(z)}::e^{ikX_{-}(\bar{z})}: \nonumber \\
\label{329}
 &=&\frac{(z-\bar{z})^2}{z\bar{z}}:e^{ik[X_{+}(z)+X_{-}(\bar{z})]}:
 =\frac{(z-\bar{z})^2}{z\bar{z}}:e^{2ikX(z,\bar{z})}:.
\end{eqnarray}
Here, $K\equiv 2k$  corresponds to a momentum of an external field coupled to the noncommutative open string $X(z,\bar{z})$.  Since $K^2=(2k)^2=8$ ( we have set $2\alpha '=1$), this means that the external field is the ground-state tachyon of the closed string. The coupling constant factor 
$g(\sigma )={(z-\bar{z})^2}/{z\bar{z}}$ is proportional to $\sin^2 \sigma $. This behavior is the same as in the ordinary open string theory without the $B$ field. \\
\indent
 Finaly it still remains to check Eq.(3.9). This equation is true if vertex operators $V(z,\bar{z})$ and $V(w,\bar{w})$ are commutable with each other.  In order to see this, let us note the following equations:
\begin{equation}
\label{330}
 U_{+}(w)U_{-}(\bar{z})=\frac{(w-\bar{z})^2}{w\bar{z}}:U_{+}(w)U_{-}(\bar{z}):
  =U_{-}(\bar{z})U_{+}(w),
\end{equation}
\begin{equation}
\label{331}
 U_{\pm}(z)U_{\pm}(w)=\frac{(z-w)^2}{zw}:U_{\pm}(z)U_{\pm}(w):
  =U_{\pm}(w)U_{\pm}(z).
\end{equation}
By using these equations one can see
\begin{eqnarray}
 [V(z,\bar{z}), V(w,\bar{w})] &=& [U_{+}(z)U_{-}(\bar{z}), U_{+}(w)U_{-}(\bar{w})] \nonumber \\
 &=& U_{+}(z)[U_{-}(\bar{z}), U_{+}(w)]U_{-}(\bar{w}) + U_{+}(w)[U_{+}(z), U_{-}(\bar{w})]U_{-}(\bar{z}) \nonumber \\
 &=& 0.   
\end{eqnarray}
This proves Eq.(3.9).

\section{Concluding remarks}
\setcounter{equation}{0}
\indent 
 
We have drived the vertex operator (\ref{329}) which describes an emission of the closed-string tachyon out of the noncommutative open string. Such a vertex operator has been shown to exist only when the momentum of the closed-string tachyon is subject to the constraints Eqs.(3.20) and (3.22). The vertex operator has a multiplicative coupling constant given by 
$g(\sigma )=\sin^2\sigma $, $0\le \sigma \le \pi $. This behavior is the same as in the ordinary open string theory without the $B$ field. \\
\indent
  The external closed-string tachyon field has been seen to couple only with those components (for the four-dimensional example in Sec.III: $X_0$, $X_1$) of the open string coordinates, which are not coupled with the $B$ field.  The system breaks down into two dynamically independent subsystems: one consisting of the components (for the four-dimensional example in Sec.III: $X_2$, $X_3$) of $X$ coupled with the $B$ field and the other consisting of the remaining components ($X_0$, $X_1$ ) coupled with the closed-string tachyon.  This seems to be the reflection of the general feature of the string interaction that the closed string cannot be coupled with the massless vector which is a member of the open string, since there does not exist a vertex of the type "closed-closed-open". \\
\indent
     The graviton is also coupled with the noncommutative open string. In the standard weak field approximation of the gravitational field $g_{\mu \nu }(X)$, the vertex operator is given as
\begin{equation}
\label{401}
 V = \epsilon _{\mu \nu }J^\mu _{\ +}(z)U_{+}(z)U_{-}(\bar{z})J^{\nu }_{\ -}(\bar{z})
  = \epsilon _{\mu \nu }J^\mu _{\ +}(z)e^{2ikX(z,\bar{z})}J^{\nu }_{\ -}(\bar{z}),
\end{equation}
where $\epsilon _{\mu \nu }$  is a polarization tensor and required constraints are 
\begin{equation}
\label{402}
 k^2 = 0, \qquad k\bar{\beta}k = 0. 
\end{equation}
In this case there is no $\sigma $-dependent coupling factor. The form of (4.1) is the same as in Ref.[10]. \\
\indent
  For simplicity we have dealt here with the neutral-open string with opposite charges at both ends.  We remark that the whole discussion is valid also for the charged-open string with an arbitrary charge at each end [13].


\end{document}